\documentclass[11pt,twoside]{article}
\usepackage{asp2010}
\resetcounters

\begin{document}

\title{Radio and X-ray Observations of Five TeV Supernova Remnants}

\author{Wenwu Tian$^{1,2}$, Denis Leahy$^2$ 
\affil{$^1$National Astronomical Observatories of China, CAS, Beijing 100012, China\\ }
\affil{$^2$Department of Physics \& Astronomy, Uni. of Calgary, AB T2N 1N4, 
Canada}}

\begin{abstract}
We briefly summarize recent results of five TeV Supernova Remnants(SNRs) from radio and X-ray observations. We focus on remeasuring kinematic distances of 5 TeV SNRs, i.e. HESS J1731-347/SNR G353.6-0.7 (3.2 kpc), HESS J1834-087/G23.3-0.3 (also W41, 4.0 kpc), HESS J1833-105/G21.5-0.9 (4.8 kpc), HESS J1846-029/G29.7-0.3 (also Kes 75, 6.3 kpc) and TeV SNR G54.1+0.3 (6.5 kpc), and studying non-thermal X-ray emissions from two old SNRs (G353.6-0.7 and W41). These not only allow constraining the TeV SNRs basic physical properties, but also help reveal acceleration mechanisms of TeV $\gamma$-rays in the SNRs which are either related with the SNRs or the pulsar wind nebulae.   
\end{abstract}

\section{Introduction}

The TeV emission from supernova remnants (SNRs) has recently become of great interest with new TeV $\gamma$-ray observations which identify some SNRs as the source \citep{Aha06}. These observations have come primarily from the HESS, CANGAROO, MAGIC and VERITAS telescope (array). Very High Energy (VHE) $\gamma$-rays are very good tracers in localizing Galactic acceleration sites of VHE particles because VHE $\gamma$-rays' propagation is not affected by the interstellar magnetic fields in the Galaxy. Most SNRs, as one of the major TeV sources, are located on the Galactic plane. 

We have carried out follow-up studies of TeV-emitting SNRs to find basic physical parameters by employing the 1420 MHz continuum and the HI+CO line spectra data from the International Galactic Plane Survey, X-ray data from $\it{SUZAKU}$ and $\it{XMM-NEWTON}$ observations. The combination of the continuum and spectral line data often gives good constraints on the distance to the TeV SNRs presenting strong radio emissions, which are otherwise poorly known, thus allows good determination of their properties such as radius, explosion energy, size, age and local interstellar medium density. VHE $\gamma$-rays have long been believed to originate from the interactions between SNR shocks and surrounding interstellar molecular clouds. CO observations to TeV SNRs and their surrounding environment can provide direct evidence if a SNR is physically associated with molecular clouds and if the interactions are strong enough or not (e.g. particle density, high accelerating efficiency etc). X-ray observations to TeV SNRs can help answer the astrophysical nature of TeV emissions, i.e. the acceleration mechanisms, and finally reveal the intrinsic physics of Galactic cosmic rays.      
 
Using HI line emission/absorption to constrain distances to Galactic SNRs face two main challenges: Firstly, to build reliable HI absorption spectra. E.g. the spatial variation in the background HI emission over the target source region may cause spurious absorption features, and produce significant uncertainty in the measurement of the absorption spectrum. Secondly, to explain the HI absorption spectra properly and convert into reasonable distances. E.g. the current rotation curve model of the Galaxy is far from perfect (even the distance from the sun to the center of the Galaxy, R$_{0}$$\sim$ 8 kpc, contains an uncertainty of $\sim$0.5 kpc; while all kinematic distances are directly proportional to R$_{0}$); the near-far kinematic distance ambiguity in the inner Galaxy, and the kinematic distance confusion due to the velocity reversal within the Perseus arm etc.  
We have developed methods to reduce these concerns \citep*{Tia07, Lea08a}. Here we summarize application of the methods to five TeV SNRs (SNRs G353.6-0.7, W41, G21.5-0.9, Kes 75 and G54.1+0.3). In addition, we have in detail studied the nature of TeV emissions from G353.6-0.7 and W41 by using multi-band data. Three other TeV sources have X-ray pulsar wind nebula (PWN) counterparts therefore have been believed to be powered by PWN shock accelerating primary electrons by inverse-Compton emission \citep{Acc10}.

\section{Kinematic Distances from HI and $^{13}$CO spectra}
We have built HI-absorption spectra to the above five SNRs, and have carefully analyzed the reality of key absorption features in the spectra, including comparison with their respective CO emission spectra. We summarize the distance measurements for the SNRs in Table 1. Fig. 1 shows an example about G54.1+0.3 which is also called as PWN G54.1+0.3, the "close cousin" of the Crab nebula \citep{Lu02,Boc10}. The highest velocity of the absorption features is beyond the tangent point (see fig. 1), hinting a lower distance limit of 4.5 kpc according to a measured rotation curve model. There is no visible absorption feature at 48 km s$^{-1}$ although there appears a bright HI emission feature at the velocity. \citet{Lea08} suggested a distance of $\sim$ 6.2 kpc based on HI absorption features and a possible morphological association between the nebula and a CO molecular cloud at velocity of $\sim$-53 km s$^{-1}$. However, the association is not solid until further supporting evidences, e.g., detection of enhanced X-ray thermal emission or  1720 MHz OH masers which are seen as signposts of the SNR-cloud interaction. New HI channel maps confirm that there are obvious HI depressions at 47 and 49 km s$^{-1}$ but no depression at 48 km s$^{-1}$ at the location of G54.1+0.3 (Fig. 1). This supports the bright HI emission at 48 km s$^{-1}$ is likely at the far side of this velocity which hints that G54.1+0.3 is more likely at a distance of $\sim$6.5 kpc. The kinematic distances in Table 1 are estimated from the measured rotation curve which generally causes an uncertainty of $\le$ 0.5 kpc \citep{Gom06}. 
\begin{table}
\begin{center}
\caption{Summary of Distances of Five SNRs}
\setlength{\tabcolsep}{1mm}
\begin{tabular}{cccccc}
\hline
SNR names: & G353.6-0.7 & G21.5-0.9 & H23.3-0.3 & Kes 75 & G54.1+0.3 \\
\hline
Highest absorption \it{v} (km/s):  & -20 & 67 & 78 & 95& 65 \\
Kinematic distance (kpc):  & $\sim$ 3.2 &  $\sim$ 4.8  &  $\sim$ 4.2&  $\sim$ 6.0 &  $\sim$ 6.5 \\
\hline
\hline
\end{tabular}
\end{center}
\end{table}

\section{X-ray Emissions of Two TeV SNRs}
Both SNRs G353.6-0.7 and W41 are old (10$^{4}-10^{5}$ yrs) but associated with TeV $\gamma$-ray sources, which is unusual \citep{Yam06, But09}. Based on theories and observations, both PWNe and young SNRs are accepted as major sources of TeV emissions because their strong shocks may be able to accelerate particles to relativistic energy and generate VHE emissions. We have carried out studies aimed at the two SNRs by using $\it{XMM-NEWTON, SUZAKU, ROSAT}$ X-ray data and CO data in order to constrain the TeV SNRs basic properties and reveal acceleration mechanisms of TeV $\gamma$-rays in old SNRs. We have found that there are likely non-thermal X-rays from the regions overlapping the HESS sources for both SNRs, and giant molecular clouds are detected overlapping the SNRs too. In addition we have detected several X-ray compact sources within the TeV sources but no pulsation signals. In summary, we suggest that these TeV $\gamma$-ray emissions originate from hadronic particles' radiations when these hadrons have been accelerated by the old SNR shocking the adjacent dense clouds then generate Pion-decay $\gamma$-rays. Here we show an example in Figure 2; more details can be found in recent works \citep{Tia10, Hal10}.      
\begin{figure*}
\vspace{90mm}
\begin{picture}(80,80)
\put(-20,170){\includegraphics{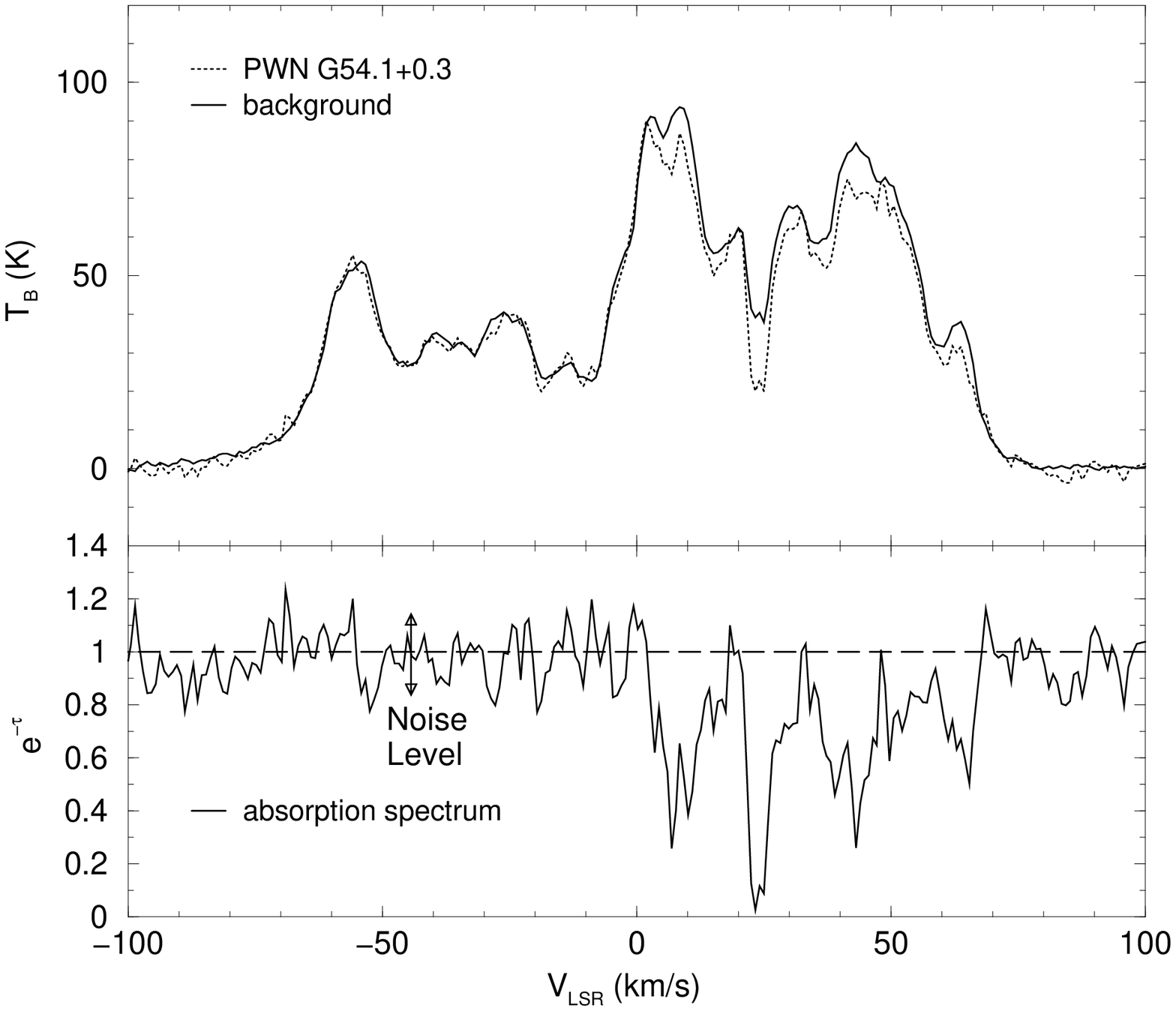}}
\put(190,110){\includegraphics{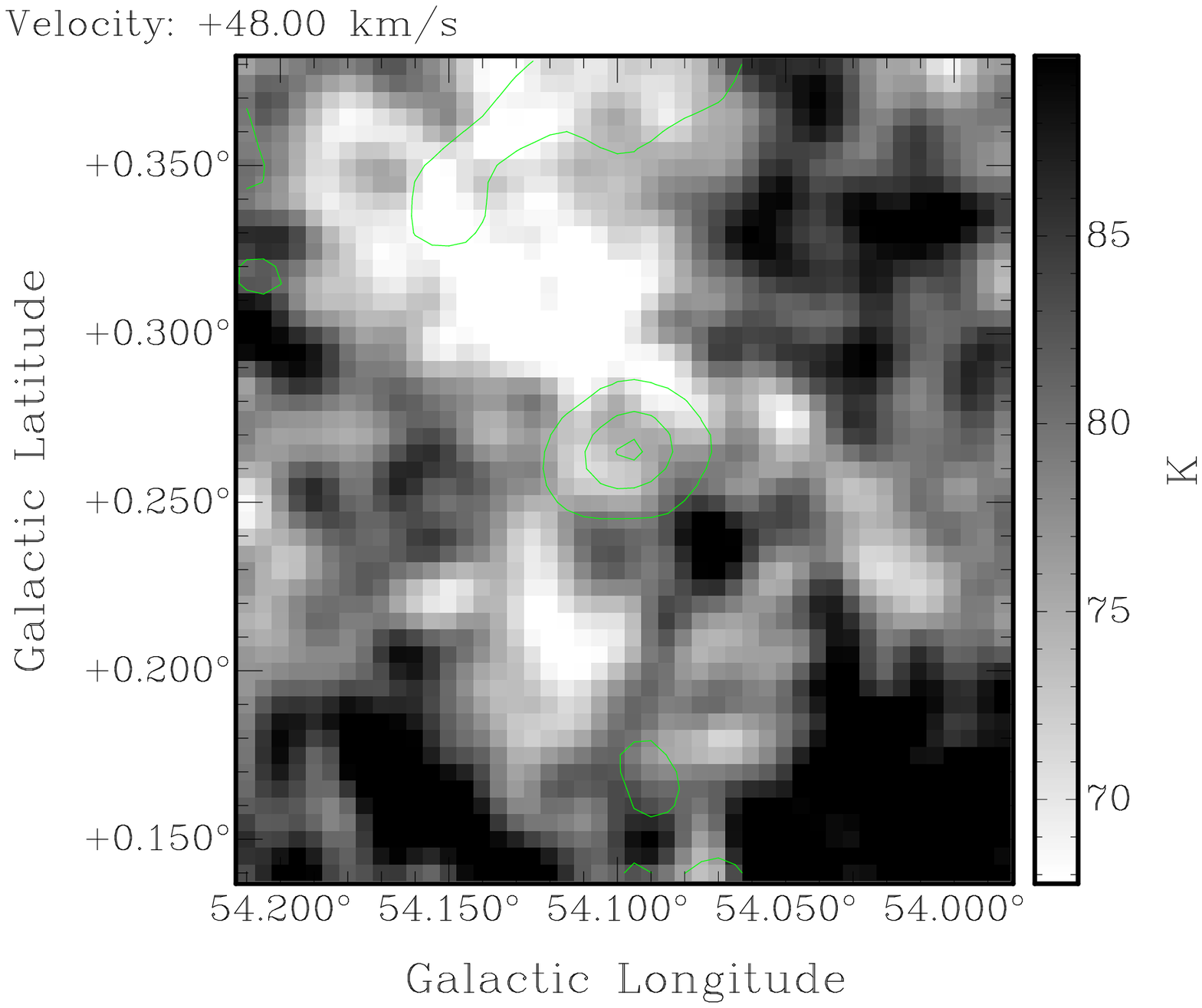}}
\put(-10,-70){\includegraphics{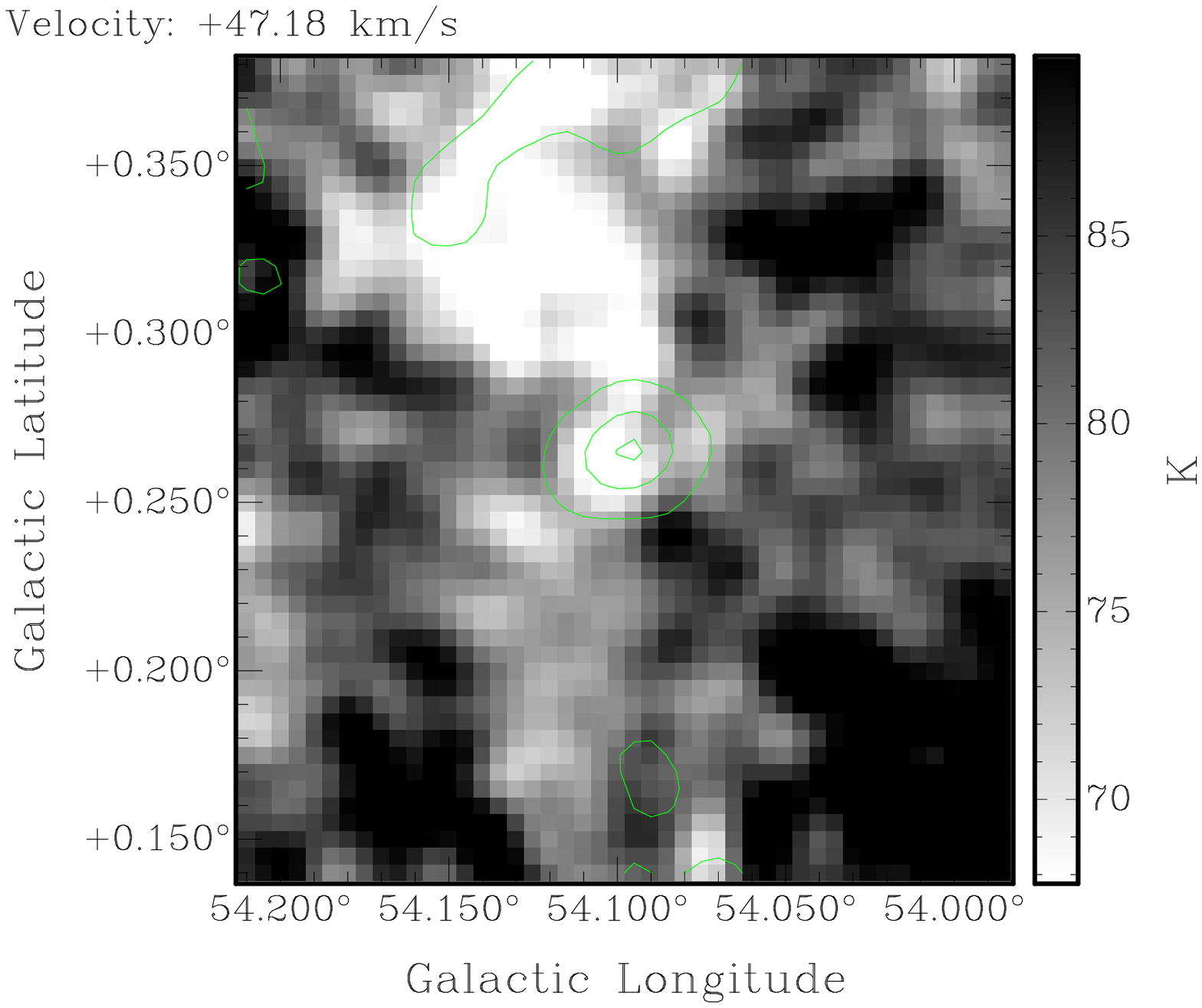}}
\put(190,-70){\includegraphics{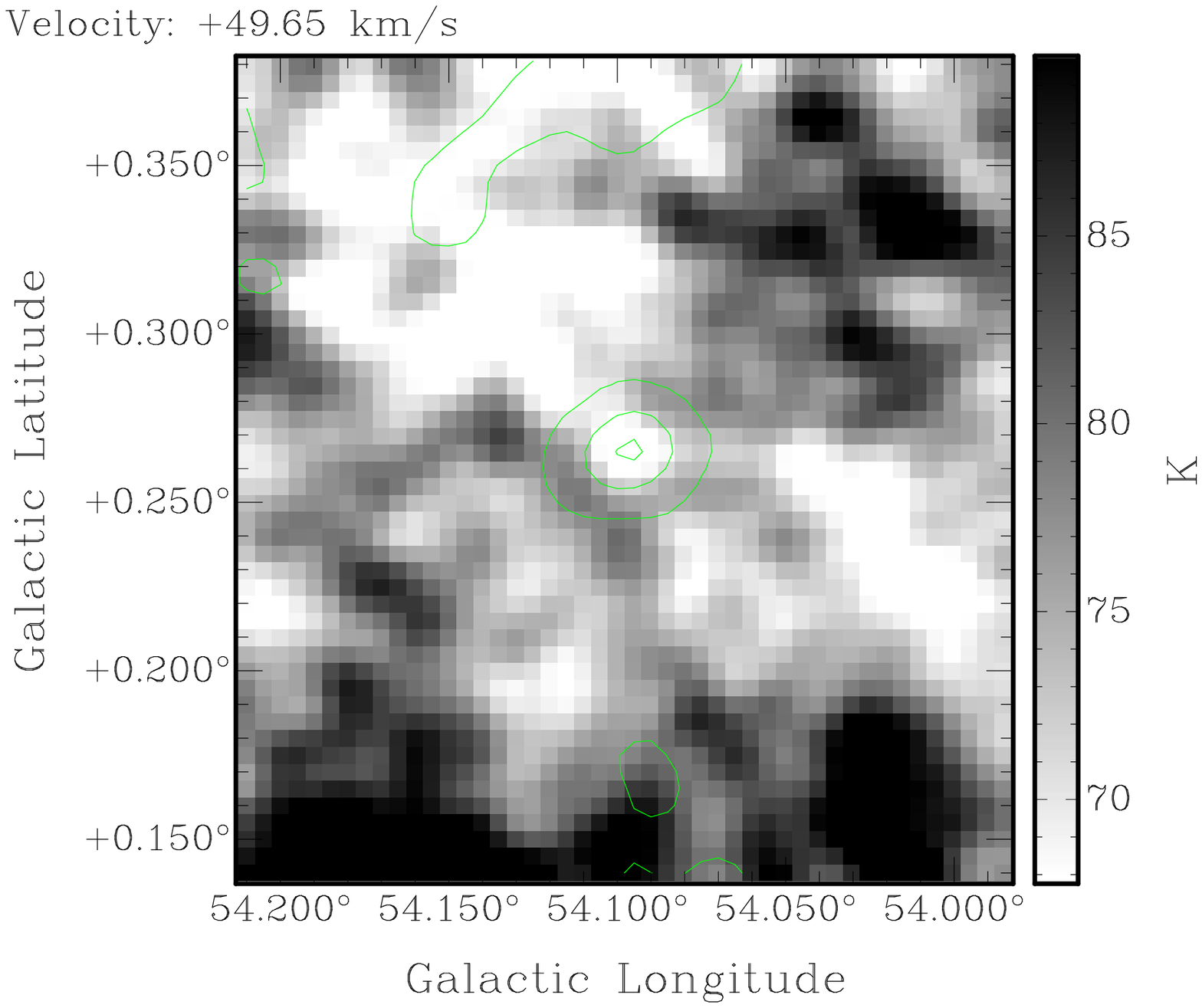}}
\end{picture}
\caption{The HI absorption spectra of SNR 54.1+0.3 (upper left), and three HI channel maps at 47, 48 and 49 km s$^{-1}$ respectively, centered at G54.1+0.3 (contours).}
\end{figure*}

\begin{figure}
\vspace{35mm} \begin{picture}(80,80)
\put(280,-20){\includegraphics{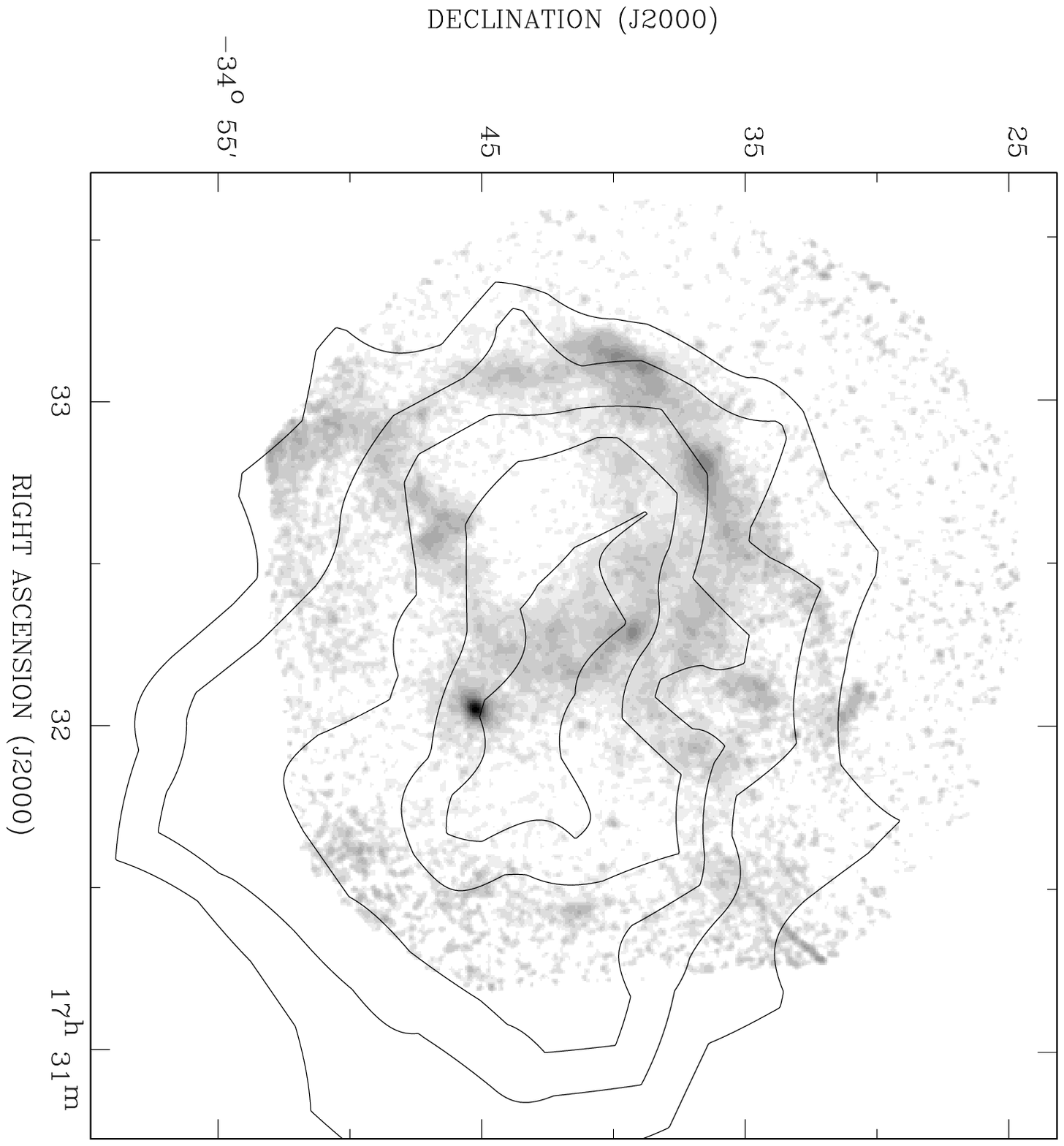}}
\put(160,-10){\includegraphics{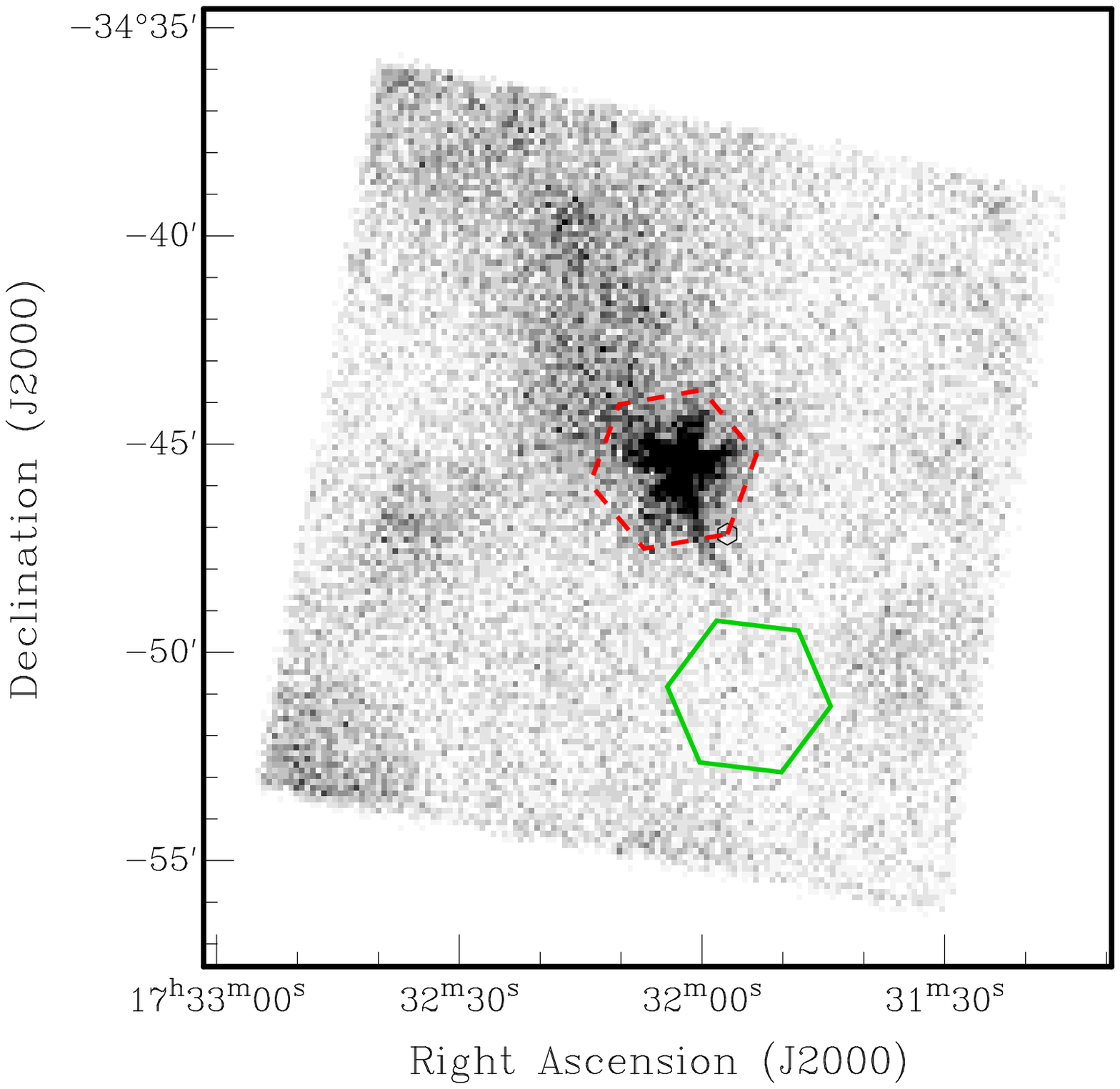}}
\end{picture}
\caption[xx]{Left: {\it{XMM-NEWTON}} 0.8-7 keV intensity image (greyscale) overlaid with contours of the TeV $\gamma$-ray emission.  
The X-rays from the shell of the SNR are likely non-thermal because its spectra have a photon spectral index of $\sim$ 2.   
Right: SUZAKU image of the central region of G353.6-0.7. The compact source's spectrum has been extracted from the region enclosed by the dash-line diamond after subtracting the background (the solid-line diamond). The spectrum's analysis of the compact source excludes the possibility that it is a PWN}.
\end{figure}

\acknowledgements WWT acknowledges supports from the Bairen project of the CAS and the NSFC. DAL thanks the NSERCC for support. 

\bibliography{tian_wenwu}

\end{document}